\documentclass[journal]{IEEEtran}
\usepackage{amsmath,amsfonts}
\usepackage{algorithmic}
\usepackage{algorithm}
\usepackage{array}
\usepackage{textcomp}
\usepackage{stfloats}
\usepackage{url}
\usepackage{enumitem}
\usepackage{verbatim}
\usepackage{graphicx}
\usepackage{cite}
\usepackage{stfloats}
\usepackage{balance} 
\usepackage{subfig}
\begin{document}
\title{Towards a Molecular Computer: Enabling
Arithmetic Operations in Molecular Communication}
\author{Jianqiao Long, Lei Zhang, Miaowen Wen, \IEEEmembership{Senior Member,~IEEE}, Kezhi Wang, \IEEEmembership{Senior Member,~IEEE}, Natalio Krasnogor, and Jichun Li \IEEEmembership{Member,~IEEE}
\thanks{ }
\thanks{Jianqiao Long, Natalio Krasnogor and Jichun Li are with the Interdisciplinary Computing and Complex BioSystems research group, School of Computing, Newcastle University, Newcastle upon Tyne, NE4 5TG, U.K. (email: j.long5@newcastle.ac.uk; natalio.krasnogor@newcastle.ac.uk; jichun.li@newcastle.ac.uk.) }
\thanks{Lei Zhang and Miaowen Wen are with the School of
Electronic and Information Engineering, South China University of Technology, Guangzhou 510641, China (email: 202320111103@mail.scut.edu.cn; eemwwen@scut.edu.cn).}
\thanks{Kezhi Wang is with Department of Computer Science, Brunel University of
London, Uxbridge, Middlesex, UB8 3PH (email: kezhi.wang@brunel.ac.uk).}}

\maketitle

\begin{abstract}
In current molecular communication (MC) systems, performing computational operations at the nanoscale remains challenging, restricting their applicability in complex scenarios such as adaptive biochemical control and advanced nanoscale sensing. To overcome this challenge, this paper proposes a novel framework that seamlessly integrates computation into the molecular communication process. The system enables arithmetic operations, namely addition, subtraction, multiplication, and division, by encoding numerical values into two types of molecules emitted by each transmitter to represent positive and negative values, respectively. Specifically, addition is achieved by transmitting non-reactive molecules, while subtraction employs reactive molecules that interact during propagation. The receiver demodulates molecular counts to directly compute the desired results. Theoretical analysis for an upper bound on the bit error rate (BER), and computational simulations confirm the system's robustness in performing complex arithmetic tasks. Compared to conventional MC methods, the proposed approach not only enables fundamental computational operations at the nanoscale but also lays the groundwork for intelligent, autonomous molecular networks.  
\end{abstract}

\begin{IEEEkeywords}
AirComp, Computation and communication cooperation, molecular communication, nanoscale.
\end{IEEEkeywords}

\section{Introduction}
\label{sec:introduction}
Molecular communication (MC) has emerged as a promising paradigm for nanoscale and bio-nano communication, employing biological processes to transmit information through chemical signals \cite{jamali2019channel}. Unlike conventional schemes that rely on electromagnetic waves, MC uses molecules as carriers, enabling communication in environments where electromagnetic methods are either inefficient or infeasible. Its biocompatibility \cite{pierobon2010physical} makes MC an attractive candidate for applications such as targeted drug delivery \cite{chude2017molecular}, industrial process control \cite{gine2009molecular}, and may open the door to next-generation solutions to environmental problems \cite{hassard2024scaling}

Over the past decade, MC research has primarily focused on improving fundamental aspects of communication. Studies have examined signal modeling, noise reduction, molecular propagation, and refined receiver designs to enhance signal reliability and transmission efficiency, mitigate inter-symbol interference (ISI), and adapt to complex propagation environments \cite{chen2021resource,jing2020power,huang2021signal}. For example, works have explored resource coordination in multi-user scenarios \cite{chen2021resource}, developed distance-based power control strategies to reduce ISI \cite{jing2020power}, and investigated both model-based and data-driven detection schemes to improve receiver performance under diverse noise conditions \cite{huang2021signal}. Further research has demonstrated how specialized molecular structures and arrays can facilitate communication in challenging conditions \cite{atakan2012nanoscale}. Although these efforts have significantly improved MC’s communication performance, none have addressed the integration of computational functionality, leaving a pivotal gap in realizing data processing and operations.

As MC moves toward supporting more complex applications, the absence of integrated computational functionality emerges as a critical limitation. Many envisioned scenarios, such as the Internet of bio-nano things (IoBNT), require systems to process and interpret molecular data in real-time to achieve adaptive biochemical control and advanced nanoscale sensing \cite{juwono2021envisioning, dissanayake2021exact}. For instance, autonomous health monitoring \cite{ghavami2020anomaly} and adaptive drug delivery demand immediate \cite{chude2017molecular} and data-driven adjustments that cannot be realized through communication alone. Although integrated computational capabilities are important, they have not yet been fully realized within MC frameworks. This has created a significant gap between current communication-oriented solutions and the computational requirements of next-generation nanoscale applications.

This gap is also evident in intrabody MC scenarios, where overcoming blood-tissue barriers, addressing spatial differences, and managing fluctuating molecular concentrations require real-time data processing \cite{al2022intrabody}. Similarly, IoBNT applications in vascular networks could also benefit substantially from the ability to perform arithmetic operations on molecular signals, enabling more precise molecule release strategies, dynamic adaptation to environmental variations, and effective real-time health monitoring \cite{lee2023internet}. Without embedded arithmetic operations, MC systems remain limited to basic message exchange, hindering their ability to act as intelligent, autonomous, and adaptive nanoscale agents.

To address this computational gap in MC, we draw inspiration from Over-the-Air Computation (AirComp), a technique originally developed in wireless communication to integrate computation directly into the communication process \cite{csahin2023survey}. Unlike conventional communication paradigms that treat computation and transmission as separate tasks, AirComp exploits the waveform superposition property of wireless multiple-access channels (MAC) to perform mathematical functions—such as summation, weighted averaging, and function aggregation—while the signals are still propagating. This eliminates the need for explicit decoding at the receiver, significantly reducing latency, bandwidth consumption, and energy requirements, making it particularly advantageous for large-scale distributed systems \cite{mitsiou2023accelerating, lin2022distributed}.

By leveraging the analog nature of wireless signals, AirComp transforms interference from a limiting factor into a computational resource, effectively turning the wireless medium into a distributed computing platform \cite{razavikia2023channelcomp, xu2021learning}. This characteristic has made AirComp a crucial enabler for real-time edge intelligence, wireless federated learning, and large-scale sensor fusion, where low-latency and scalable computation are essential \cite{zhu2018mimo, fang2021over}. Moreover, the introduction of digital AirComp techniques has expanded its applicability to modern digital communication networks, improving robustness and compatibility with next-generation wireless standards\cite{qiao2024massive, you2023broadband}.

Since its inception, AirComp has rapidly evolved from basic analog function computation to a sophisticated enabler for distributed optimization, multimodal sensing, and AI-driven communication networks\cite{jiang2021joint, csahin2023survey}. In federated learning, AirComp enables direct aggregation of local model updates from distributed devices, significantly reducing the communication overhead of large-scale machine learning frameworks \cite{jing2022federated}. Similarly, in massive Internet of Things (IoT) applications, AirComp facilitates real-time wireless data aggregation, allowing thousands of edge devices to compute global statistics while minimizing transmission delays and energy costs \cite{zhu2021over}. Additionally, broadband AirComp has been explored to enable reliable AirComp in vehicular networks, drone swarms, and next-generation mobile edge computing systems \cite{wen2023task}.

Given its ability to seamlessly integrate communication and computation, AirComp has transformed wireless networks from mere data transmission systems into intelligent, decentralized computing platforms. This paradigm shift provides valuable insights for enabling computational functionality in MC. By leveraging the principles of AirComp, we can develop novel MC frameworks that perform real-time molecular arithmetic and adaptive biochemical processing during signal propagation, paving the way for autonomous nanoscale decision-making in distributed molecular networks.

Integrating computation into MC would constitute a transformative advancement, allowing MC systems to evolve beyond mere communication links into nanoscale computing platforms. Such platforms could perform arithmetic operations, both fundamental and advanced, directly on molecular signals during propagation. This integrated computational capability would facilitate advanced functionalities—ranging from adaptive diagnostics and targeted therapeutics to the implementation of rudimentary nanoscale machine learning algorithms—all executed directly within the communication channel.

In our previous work \cite{long2024computation}, we took an initial step toward this goal by demonstrating the feasibility of addition in MC systems. However, the inability to perform other fundamental arithmetic operations (e.g., subtraction, multiplication, and division) continues to limit the computational expressiveness and adaptability of MC. Building on our earlier findings, this paper proposes a comprehensive MC framework that extends computation to include all four fundamental arithmetic operations. We represent numerical values as molecular concentrations, one for positive values and the other for negative values. When multiple transmitters emit the same species simultaneously, their concentrations aggregate at the receiver, enabling addition. Conversely, if opposing species are released together, controlled chemical reactions occur during propagation, effectively subtracting their concentrations. As a result, the receiver can directly measure the final molecular levels to obtain the computed arithmetic results.

By enabling such integrated computation, this work significantly broadens the application scope of MC, laying the groundwork for nanoscale communication networks that can execute complex, real-time computations.

The contributions of this work are as follows: 
\begin{itemize} 
   \item We propose a new molecular communication framework capable of performing arithmetic operations during molecular propagation, thereby enhancing the computational functionality of MC systems. By leveraging AirComp principles, our approach unifies communication and computation, facilitating advanced bio-nano applications.
   \item An upper bound for the bit error rate (BER) is derived to evaluate the feasibility of integrated computational operations, incorporating the combined effects of diffusion-driven noise and reaction-based interactions to provide a reliable basis for performance assessment in practical molecular communication scenarios.
   \item The proposed framework is validated through simulations, demonstrating high accuracy and practicality in noisy environments for bio-nano applications. Under diverse channel conditions, it consistently retains a relatively high detection accuracy, underscoring its viability for real-world deployments.
\end{itemize}

The rest of this paper is organized as follows. Section II presents the system model, including the framework for integrating fundamental arithmetic operations in MC. Section III describes a detailed theoretical analysis, including the derivation of the BER upper bound. Section IV discusses the simulation setup and results that validate the effectiveness of the proposed framework. Finally, Section V concludes the paper and outlines future research.

\begin{figure}[ht]
    \centering
    \subfloat[Illustration of the basic principle for a summation MC system.]{
        \includegraphics[width=1.0\linewidth]{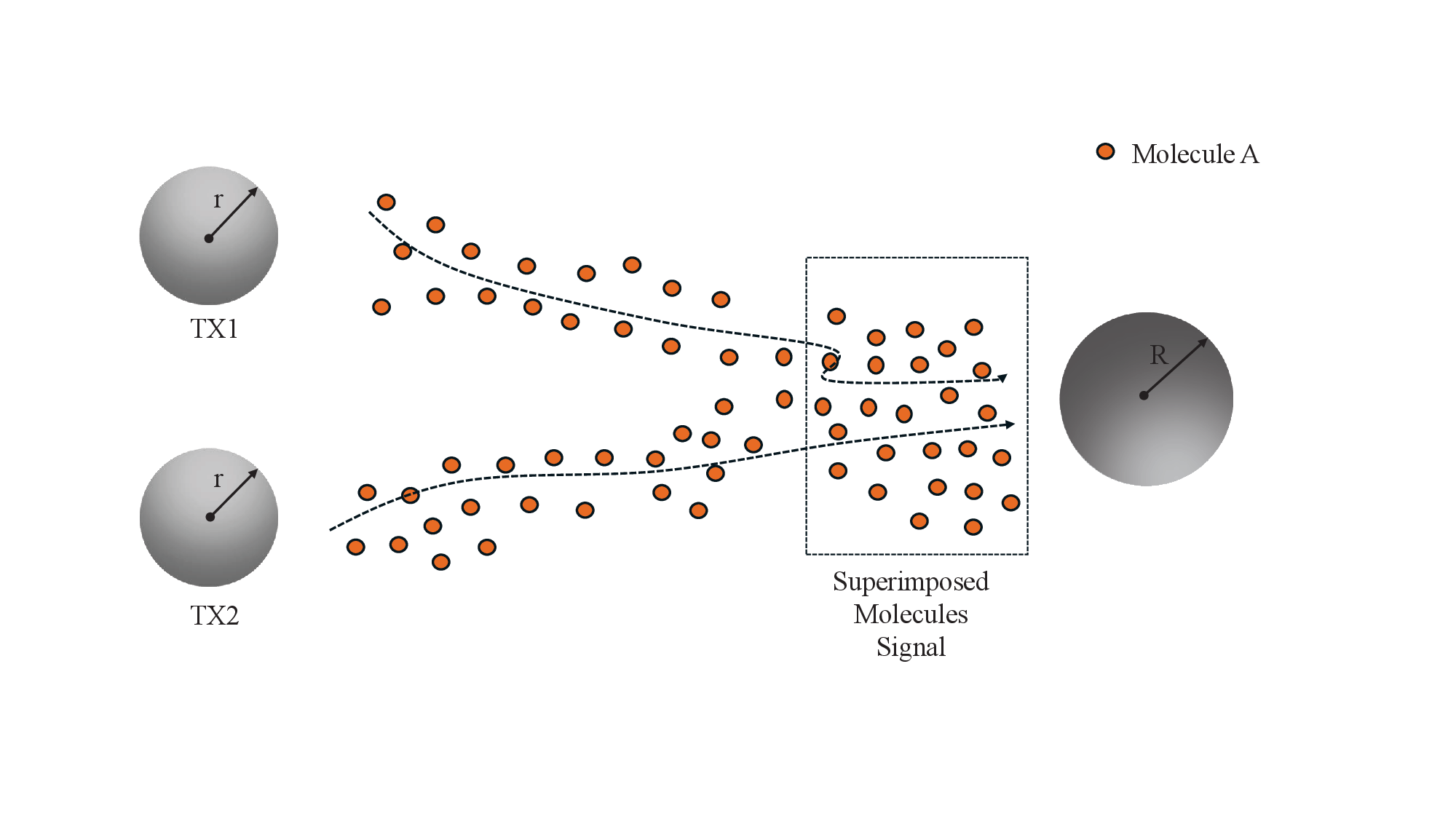}
        \label{fig:sub_aa}
    }
    \hfill
    \subfloat[Illustration of the basic principle for a subtraction MC system.]{
        \includegraphics[width=1.0\linewidth]{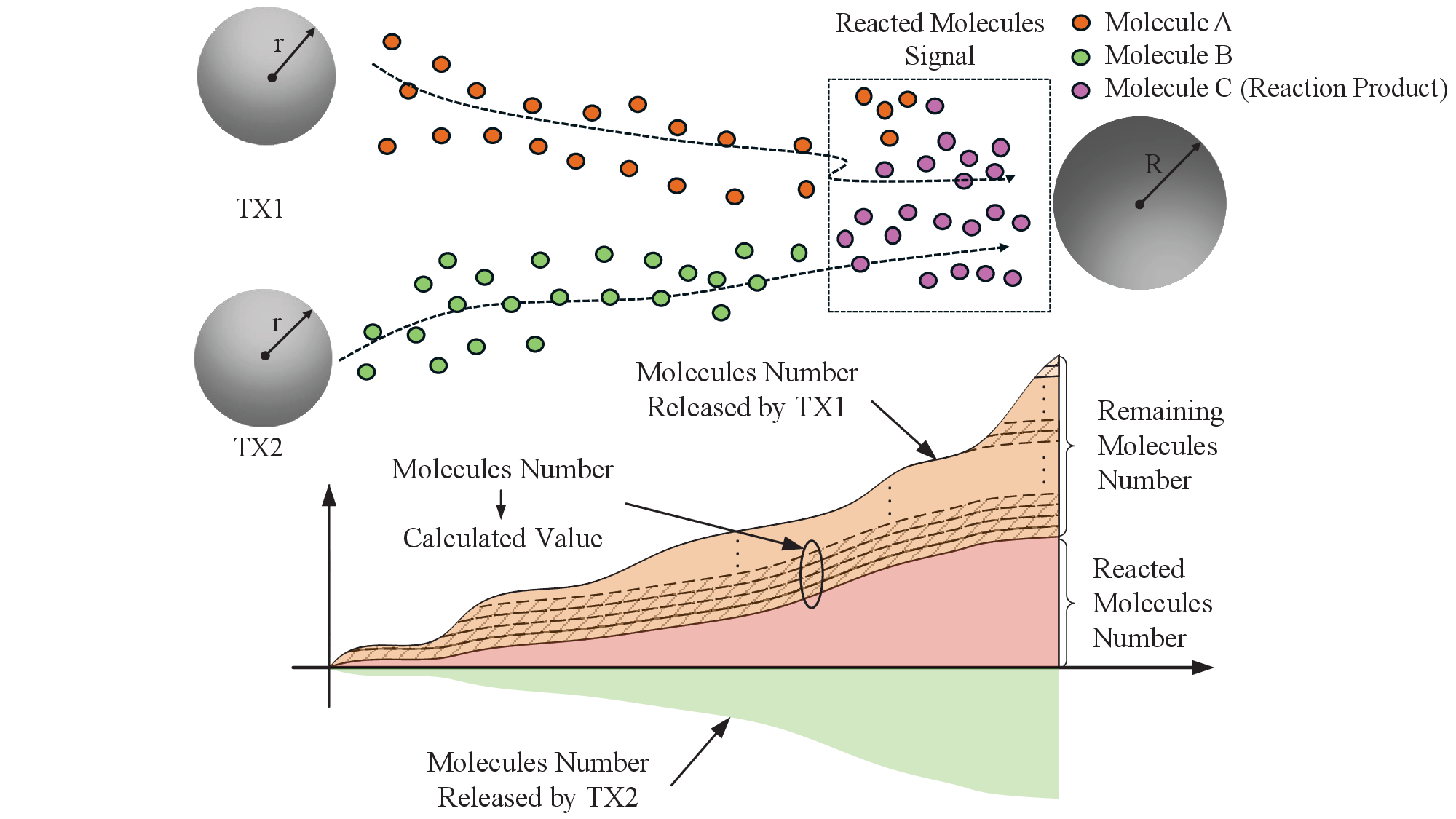}
        \label{fig:sub_bb}
    }
    \caption{Two paradigms: (a) An MC system with a summation computation function; (b) An MC system with a subtraction computation function.}
    \label{fig:ccccc}
\end{figure}

\section{system model}

This section describes the proposed MC system, which integrates arithmetic computation directly into the communication process. By controlling the emission of two molecular species, one for positive and the other for negative values, the system enables arithmetic operations: identical species aggregate to perform summation, while opposing species neutralize each other to achieve subtraction. The following subsections present an overview of the system, transmitter and receiver models, and the bit transmission framework.

\subsection{Overview of the Model}
The proposed MC system utilizes multiple transmitters and a single receiver to achieve computations via molecular interactions during signal propagation. Fig.~\ref{fig:ccccc} illustrates the fundamental principles and operational modes of the system.

In Fig.~\ref{fig:ccccc}(a), the system performs addition directly within the communication channel. Two transmitters (TX1 and TX2) emit positive molecule $A$, which propagates and naturally aggregates. At the receiver, the resulting concentration inherently represents the sum of the transmitted values. Summation can be achieved with either of  Molecules A (lets say to sum possitive values) or Molecules B (to sum negative ones). By embedding summation into the propagation process, the system overcomes the inherent difficulties of performing arithmetic operations at the nanoscale, thereby achieving the integration of computation and communication.

In Fig.~\ref{fig:ccccc}(b), the molecular communication system implements a subtraction operation through controlled chemical reactions occurring during propagation. The transmitters, TX1 and TX2, release positive molecule $A$ and negative molecule $B$, respectively. As these molecules diffuse through the medium, they undergo an `` interaction". In our model, an interaction between molecules $A$ and $B$ might be embodied by one of the following processes:
\begin{enumerate}[label=(\alph*)]
    \item $A$ degrades $B$.
    \item $A$ binds to $B$ (sequestering or complexifying with it).
    \item $A$ blocks (some of) the receptors on $R$ that would otherwise sense and count $B$.
    \item $A$ catalytically modifies $B$, thus potentially serving as a ``gain'' or ``scaling'' factor in the computed results.
    \item $A$ indirectly modulates (e.g., via post-translational modification) $B$.
\end{enumerate}

The overall effect of any of these processes is to effectively reduce the number of $B$ molecules ``sensed'' by the receiver as a function of the A molecules that are present. These interactions are quite ubiquitous in nature. For example, matrix metalloproteinases ($A$) secreted by certain cells degrade components of the extracellular matrix, such as a signaling peptide $B$ that a receptor cell senses for downstream processes; in an antibody--antigen pair, $A$ is an antibody that binds $B$ (the antigen), forming a complex that modulates the free concentration of $B$ reaching a receptor or responder cell $R$; competitive inhibition in neurotransmission provides another good analogy, e.g., naloxone ($A$) binds to the mu-opioid receptors in brain cells, blocking the binding of opioid ligands ($B$), so that $A$ acts like an antagonist occupying $B$’s receptors on neurons or immune cells, thus reducing $B$’s signaling (and hence counts); prohormone convertases process inactive precursors into active hormones, where $A$, acting as an enzymatic catalyst, converts $B$ from a precursor form to an active form, thereby modulating the effective count level of $B$; and finally, phosphorylation is a widespread process by which a receptor’s $R$ response to its cognate molecule’s concentration is indirectly modulated. Moreover, processes such as those in (a)–(e) are readily implementable using synthetic genetic circuits~\cite{konur2021toward}, sophisticated DNA strand displacement systems (e.g., as in~\cite{lopiccolo2021last}), and a variety of other chemistries~\cite{fellermann2014formalizing}.
Assuming a 1:1 reaction ratio between $A$ and $B$, the number of remaining molecules at the receiver provides direct information about the transmitted difference, which can be ``read out”, e.g., as the concentration of Moleculas $A$ (or $B$) at the receiver.

The graphical representation in Fig.~\ref{fig:ccccc}(b) illustrates the molecular subtraction process through area-based visualization. The yellow-shaded region represents the total number of molecules released by TX1 (molecule $A$), while the green-shaded region corresponds to the molecules released by TX2 (molecule $B$). Since TX2's molecules represent a negative quantity in the subtraction process, this region is positioned below the reference axis. The red-shaded area indicates the number of molecules that have reacted to form molecule $C$, verifying the occurrence of chemical interactions.

At the receiver, the subtraction result is derived by computing the difference between the yellow and green areas, which corresponds to the number of unreacted molecules detected. The remaining molecules, represented by the dashed lines, are mapped to the calculated subtraction value. Multiple dashed lines indicate different potential output values based on varying transmission conditions. Additionally, the purple-colored region highlights the reacted molecules (molecule $C$), providing further validation of the underlying reaction mechanism. This approach enables a direct and real-time subtraction operation in molecular communication, leveraging reaction-diffusion dynamics to perform arithmetic computations in biochemical environments.

\subsection{Transmitter Model}
The transmitter (TX) is responsible for encoding numerical values into molecular signals, enabling arithmetic operations directly within the communication channel. Each TX emits two types of molecules to represent positive and negative values, facilitating addition and subtraction operations. Molecules of type $A$ encode positive values, accumulating together during diffusion to represent a sum, while molecules of type $B$ encode negative values and chemically react with any $A$ molecules they encounter, effectively computing the difference as the reaction proceeds. Molecules are released in quantities proportional to the magnitude of the encoded value, ensuring that the molecular concentration accurately reflects the data being transmitted. To prevent interference and maintain computational accuracy, all transmitters are synchronized to emit molecules within designated intervals. This synchronization ensures that molecular interactions occur as intended. 

\subsection{Channel and Receiver Model}

We consider a conventional single-transmitter--single-receiver MC system in a three-dimensional, unbounded fluid medium, where molecules diffuse freely via Brownian motion at a constant diffusion coefficient. At the initial time, the transmitter (Tx) releases $Q$ molecules into the medium, which diffuse according to Fick’s second law:
\begin{align}
    \label{eq:FicksLaw}
    \frac{\partial \mathcal{C}(x,y,z,t)}{\partial t} 
    = D \nabla^2 \mathcal{C}(x,y,z,t),
\end{align}
where $\mathcal{C}(x,y,z,t)$ represents the molecule concentration at location $(x,y,z)$ and time $t$, and $D$ denotes the diffusion coefficient of the medium. Solving this diffusion equation yields the molecule concentration at the receiver:
\begin{align}
    \label{eq:Concentration}
    \mathcal{C}(t) 
    = Q \frac{V}{\bigl(4\pi D t\bigr)^{3/2}} 
      \exp\!\Bigl(-\tfrac{d^{2}}{4Dt}\Bigr),
\end{align}
where $V = \tfrac{4}{3}\pi r^3$ is the receiver volume, $d$ is the distance between the transmitter and receiver centers, and $r$ is the receiver radius. From this result, the channel impulse response (CIR), i.e., the probability that a single molecule is observed at the receiver, is given by:
\begin{align}
    \label{eq:CIR}
    h(t) 
    = \frac{V}{(4\pi D t)^{3/2}} 
      \exp\!\Bigl(-\tfrac{d^2}{4Dt}\Bigr).
\end{align}

Within this MC system, the channel supports both aggregation and reactive interactions among molecules, enabling arithmetic operations---such as addition and subtraction---directly during propagation. The receiver (RX) is modeled as a transparent spherical boundary, allowing molecules to pass freely while accurately counting the number of molecules within its volume. For summation, the RX measures the concentration of non-reactive molecular species (representing positive values) that naturally accumulate during diffusion, thus reflecting the total sum of the transmitted inputs. Conversely, subtraction is realized by having opposing molecular species interacting during propagation, leaving only the remaining non-interacting molecules at the RX to indicate the difference. This capability to directly interpret molecular interactions in the channel obviates the need for dedicated computational hardware or wetware at the receiver, thereby reducing system latency improving operational efficiency 
and simpliflying potential system implementations for arithmetic tasks in bio-nano applications.

\subsection{Bit Transmission and Operation}
In the proposed system, numerical values are represented through controlled molecular concentrations, with two distinct molecular species designated to encode the sign: one for positive values and the other for negative values. To represent an integer \( x \), the transmitter releases a number of molecules proportional to \( |x| \), choosing the positive species if \( x > 0 \) and the negative species if \( x < 0 \). By adjusting the number of emitted molecules, a wide range of integers, such as 1, 2, 3, or 4, can be encoded within each transmission interval.

Arithmetic operations arise naturally from the molecular interactions within the propagation environment. When identical, the same non-reactive molecules diffuse and accumulate, their combined concentration at the receiver represents the sum of the transmitted values. Conversely, when opposing species interact, they neutralize one another, effectively performing subtraction. By repeating addition operations, the system achieves multiplication, while repeated subtraction operations enable division, thereby extending its capabilities to all four fundamental arithmetic operations.

At the receiver, measuring the resulting molecular concentrations at the end of each interval directly yields computed results, integrating both data transmission and computation into a single process. 

\section{system principle and mathematical derivation}

This section elaborates on the theoretical foundation and mathematical modeling of the proposed MC system. We begin with the principles of a traditional single-transmitter, single-receiver diffusion-based MC system and extend it to our proposed computational MC model with multiple transmitters and a single receiver. The section includes derivations of the interference modeling, and the probability density functions (PDFs) governing the system dynamics.

To ensure reliable detection at the receiver, we adopt an equally spaced multi-sampling strategy within a single symbol period. Let \(I\) denote the number of sampling points, and let \(T_s\) represent the symbol duration. The received signal at the \(k\)-th symbol period and \(i\)-th sampling time is expressed as:
\begin{align}
    \label{Y}
    Y_k(i) = \sum_{j=0}^{L} Q_{k-j} h\left(\left(\frac{i}{I} + j\right)T_s\right) + n_k(i),
\end{align}
where \(L = \min\{k-1, l\}\) denotes the ISI length, \(Q_k\) is the number of molecules emitted during the \(k\)-th symbol period, and \(n_k(i)\) represents noise. The received signal can be decomposed as:
\begin{align}
    \hspace{-0.85em}Y_k(i) \hspace{-0.2em}=\hspace{-0.2em} \underbrace{Q_k h\left(\frac{i}{I}T_s\right)}_{\text{desired signal}} 
    \hspace{-0.2em}+\hspace{-0.2em}\underbrace{\sum_{j=1}^{L} Q_{k-j} h\left(\left(\frac{i}{I} + j\right)T_s\right)}_{\text{ISI}} 
    +\hspace{-0.2em} \underbrace{n_k(i)}_{\text{noise}}. 
\end{align}

Thus, the total received signal comprises the desired signal component \(\lambda_d\), the ISI component \(\lambda_{\text{ISI}}\), and the noise term \(n_k(i)\).

Given that the CIR \(h(t)\) represents a low probability of a single molecule arriving at the receiver, and assuming a sufficiently large number of emitted molecules, the received signal can be modeled as a Poisson process. The probability density function (PDF) at the \(k\)-th symbol period and \(i\)-th sampling time is given by:
\begin{align}
    P(Y_k(i) = N(i)) = \frac{\Lambda(i)^{N(i)} e^{-\Lambda(i)}}{N(i)!},
\end{align}
where \(N(i)\) is the observed molecule count at the \(i\)-th sampling point, and \(\Lambda(i) = \lambda_d + \lambda_{\text{ISI}}\) is the Poisson parameter.

In our proposed molecular computation system, multiple transmitters and a single receiver are considered. Each transmitter is capable of emitting two types of molecules: \(A\) for positive values and \(B\) for negative values. During signal preprocessing, type \(A\) molecules are emitted to encode addition, while type \(B\) molecules represent subtraction. Upon contact, molecules \(A\) and \(B\) react and neutralize each other. At the receiver, the observed molecule type determines the computation result: a predominance of \(A\) molecules indicates a positive outcome, while \(B\) molecules indicate a negative result. Notably, this system can also process purely additive signals. For scenarios where multiple transmitters are placed on the same side of the receiver (e.g., within a blood vessel), the received signal for the subtraction model can be expressed as the difference in probability densities for \(A\) and \(B\) molecules:
\begin{align}
    P(Y_k(i) = N(i)) = \frac{\Lambda_A(i)^{N_A(i)} e^{-\Lambda_A(i)}}{N_A(i)!} \nonumber \\
    - \frac{\Lambda_B(i)^{N_B(i)} e^{-\Lambda_B(i)}}{N_B(i)!}.
\end{align}
Here, \(\Lambda_A(i)\) and \(\Lambda_B(i)\) represent the Poisson parameters for molecules \(A\) and \(B\), respectively, determined by their diffusion coefficients and emission quantities. Multiple transmitters emitting the same molecule type contribute cumulatively to the Poisson distribution, and the sum of independent Poisson processes remains Poisson distributed. The observed molecule count \(N(i) = N_A(i) - N_B(i)\) reflects the net result, where the sign indicates the dominant molecule type, and the magnitude represents the molecular difference.

The resulting distribution follows a Skellam distribution, which incorporates a Bessel function. However, for large parameter values, the Bessel function becomes computationally unstable. To mitigate this, we approximate the Poisson distributions with Gaussian distributions:
\begin{align}
\hspace{-1.0em}P(Y_k(i) = N(i)) = \frac{1}{\sqrt{2\pi \sigma^2(i)}} 
\exp\left(-\frac{(N(i) - \mu(i))^2}{2\sigma(i)^2}\right),
\end{align}
where \(\mu(i) = \Lambda_A(i) - \Lambda_B(i)\), and \(\sigma(i)^2 = \Lambda_A(i) + \Lambda_B(i)\).

At the receiver, a multi-point MAP demodulation scheme is employed:
\begin{align}
    \hat{\mathbf{R}} = \arg \max_{\mathbf{R}_r \in \mathcal{S}} \sum_{i=1}^I \ln P(Y_r(i) = N(i) | \mathbf{R}_r),
\end{align}
which simplifies to:
\begin{align}
    \hat{\mathbf{R}} = \arg \max_{\mathbf{R}_r \in \mathcal{S}}\sum_{i=1}^I \Bigg[ 
    - \frac{(N(i) - \mu(i))^2}{2\sigma(i)^2} 
    - \frac{1}{2} \ln(2\pi \sigma(i)^2) \Big| \mathbf{R}_r \Bigg].
\end{align}
Here, \(\mathcal{S}\) represents all possible computation combinations. If there are \(M\) transmitters and \(D\) discrete computation states per transmitter, the size of the computation set is \(|\mathcal{S}| = D^M\).

\section{performance analysis}
In this section, we analyze the BER performance for the proposed system.

The MAP detection aims to maximize the posterior probability of the transmitted symbol \(R_r\) given the observed molecular counts \(N(i)\) at \(M\) sampling points. An error occurs if an alternative hypothesis \(R_e\) provides a higher likelihood. The MAP decision rule is expressed as:
\begin{align}
\sum_{i=1}^M \Bigg[ 
    -\frac{(N(i) - \mu_r(i))^2}{2\sigma_r(i)^2} 
    - \frac{1}{2} \ln (2\pi \sigma_r(i)^2) \Big| R_r 
\Bigg] \nonumber \\
\leq 
\sum_{i=1}^M \Bigg[ 
    -\frac{(N(i) - \mu_e(i))^2}{2\sigma_e(i)^2} 
    - \frac{1}{2} \ln (2\pi \sigma_e(i)^2) \Big| R_e 
\Bigg],
\end{align}
where \(N(i)\) is the observed molecular count at sampling point \(i\), \(\mu_r(i)\) and \(\mu_e(i)\) are the expected molecular counts under \(R_r\) and \(R_e\), respectively, and \(\sigma_r(i)^2\), \(\sigma_e(i)^2\) are the corresponding variances. Rearranging terms, the decision condition is reformulated as:
\begin{align}
\sum_{i=1}^M \Bigg[ 
    \frac{(N(i) - \mu_e(i))^2}{2\sigma_e(i)^2} 
    - \frac{(N(i) - \mu_r(i))^2}{2\sigma_r(i)^2} 
    \nonumber \\ + \frac{1}{2} \ln \left(\frac{\sigma_e(i)^2}{\sigma_r(i)^2}\right) 
\Bigg] 
\leq 0.
\end{align}

The decision metric \(Z\) is defined as:
\begin{align}
Z = \sum_{i=1}^M \Bigg[ 
    \frac{(N(i) - \mu_e(i))^2}{2\sigma_e(i)^2} 
    - \frac{(N(i) - \mu_r(i))^2}{2\sigma_r(i)^2} 
    \nonumber \\ + \frac{1}{2} \ln \left(\frac{\sigma_e(i)^2}{\sigma_r(i)^2}\right) 
\Bigg],
\end{align}
and an error occurs when \(Z \leq 0\). Under the Gaussian approximation, \(N(i)\) is modeled as a normal random variable:
\[
N(i) \sim \mathcal{N}(\mu_r(i), \sigma_r(i)^2) \quad
\]
If $M$ is sufficiently large, the distribution of $Z$ can be approximated by a Gaussian random variable, which allows for the derivation of its expected value and variance. Under this approximation, the expected value is given by:
\begin{align}
E[Z] = \sum_{i=1}^M \Bigg[ 
    \frac{\sigma_r(i)^2 + (\mu_r(i) - \mu_e(i))^2}{2\sigma_e(i)^2} 
    - \frac{1}{2} 
    \nonumber \\ + \frac{1}{2} \ln \left(\frac{\sigma_e(i)^2}{\sigma_r(i)^2}\right) 
\Bigg],
\end{align}
where the first term reflects the difference in likelihood contributions between hypotheses, and the logarithmic term adjusts for variance normalization. The variance of \(Z\) is:
\begin{align}
\hspace{-1em}\text{Var}(Z) \hspace{-0.2em}= \sum_{i=1}^M \Bigg[ 
    \frac{2\sigma_r(i)^4 + 4\sigma_r(i)^2 (\mu_r(i) - \mu_e(i))^2}{4\sigma_e(i)^4} 
     + \frac{1}{2} 
\Bigg].
\end{align}

The error probability corresponds to the likelihood of \(Z \leq 0\) and is expressed using the Q-function:
\begin{align}
P_\text{error} = Q\left(\frac{E[Z]}{\sqrt{\text{Var}(Z)}}\right),
\end{align}
where:
\begin{align}
Q(x) = \frac{1}{\sqrt{2\pi}} \int_x^\infty e^{-t^2/2} dt.
\end{align}
This formulation provides a theoretical upper bound for the BER, accounting for molecular noise and ISI effects, and establishes the system's error performance under varying signal and noise conditions.

\section{simulation result}
In this section, we analyze the results of the numerical simulation to verify the effectiveness of the proposed MC framework. The results include comparisons between the theoretical and simulated CIR and error rates under various configurations of the system. 
\begin{figure}[!t]
    \centering
    \includegraphics[width=1.0\linewidth]{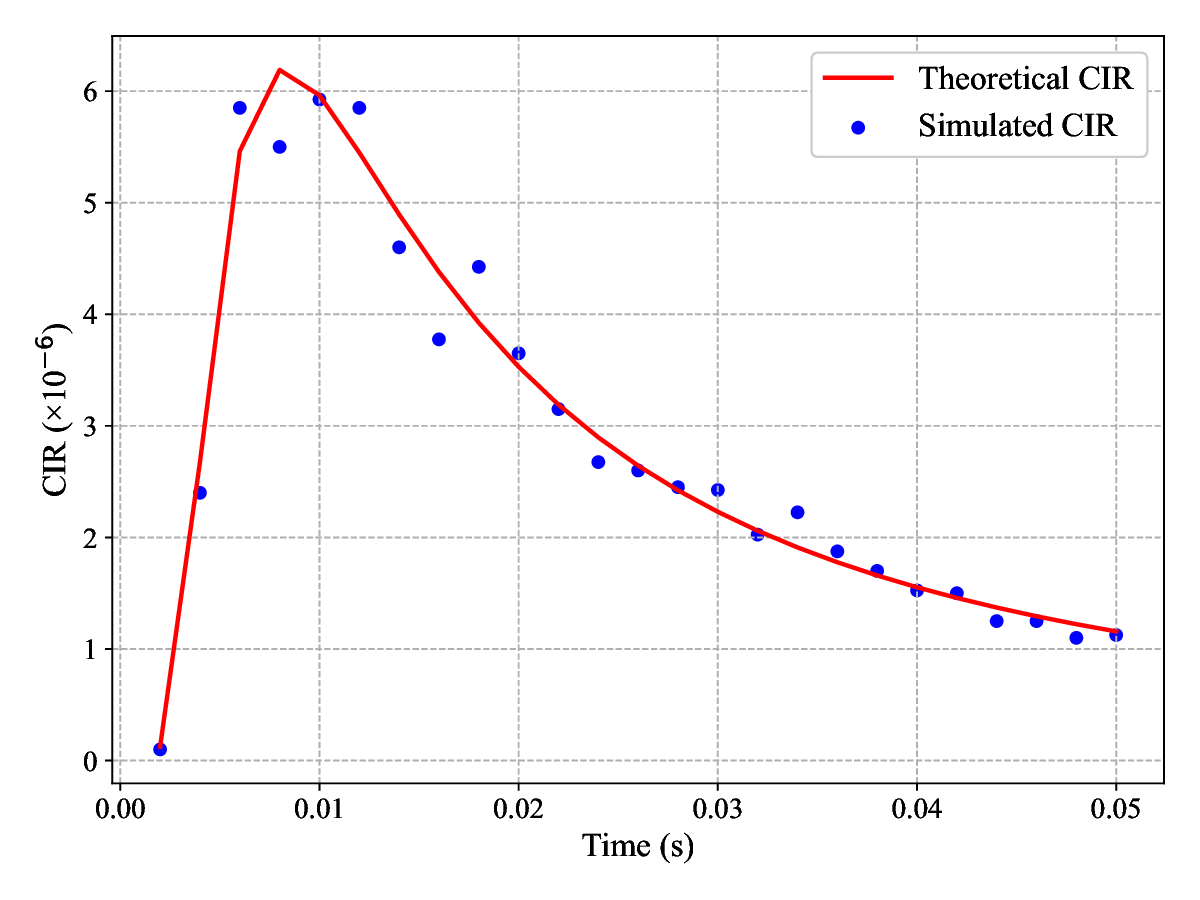} 
    \caption{Comparison of theoretical and simulated CIRs for subtraction in particle-based MC. }
    \label{fig:2}
\end{figure}
\begin{figure}[!b]
    \centering
    \includegraphics[width=1.0\linewidth]{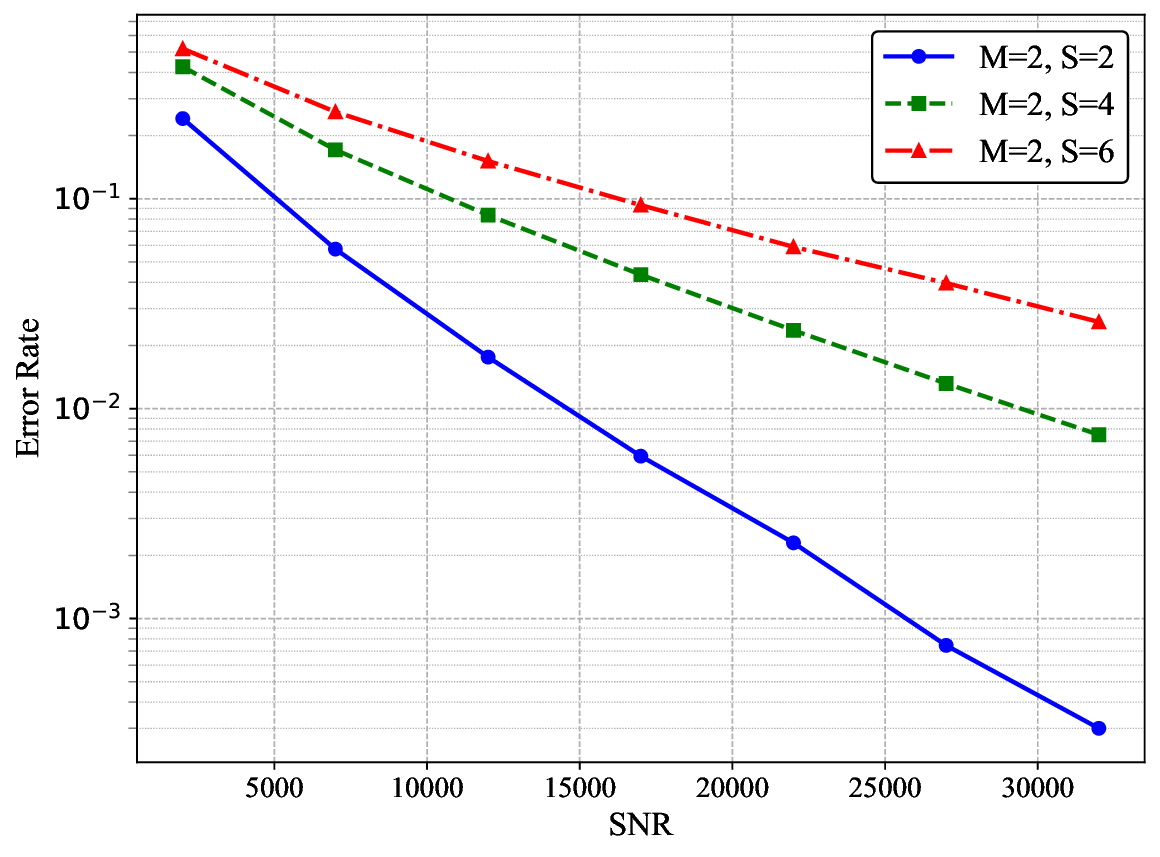} 
    \caption{Error rate performance for different computation combinations with varying SNRs. }
    \label{fig:3}
\end{figure}

We begin by examining the particle-based simulation results, Fig.~\ref{fig:2}, which depict the CIR under both theoretical and simulated models for the subtraction scenario. The theoretical CIR is derived analytically from Fick’s second law of diffusion, while the simulated CIR is obtained through Monte Carlo methods that incorporate reactive interactions between positive and negative species. The strong agreement between the theoretical and simulated CIR—particularly in the main peak and subsequent exponential decay—indicates that the theoretical model accurately captures the temporal dynamics of molecular diffusion. Minor discrepancies observed in the tail region can be attributed to the inherent randomness in molecular arrival times within particle-based simulations. Overall, the consistency between theory and simulation underscores the robustness of the proposed framework in modeling subtractive molecular diffusion processes.

Next, we analyze the system performance with varying computational combinations \( S \), as illustrated in Fig. \ref{fig:3}. We define the computational combination \( S \) as the total number of distinct absolute values that the transmitter can release. Specifically, if the transmitter is allowed to release the numerical set \(\{ -1, -2, +1, +2 \}\), then, since positive and negative values are represented by distinct molecular species, the transmitter only needs to release the corresponding set of absolute values \( \{ 1,2 \} \). Therefore, in this case, there exist 2 computational combinations. This definition generalizes accordingly for other configurations. The error rate (ER) in our system is defined as the ratio of the number of computational errors at the receiver to the total number of received signals and is expressed as:
\begin{align}
ER = \frac{N_{\text{err}}}{N_{\text{total}}},
\end{align}
where \( N_{\text{err}} \) represents the number of incorrectly received computational results, and \( N_{\text{total}} \) is the total number of received signals.

\begin{figure}[t]
    \centering
    \includegraphics[width=1.0\linewidth]{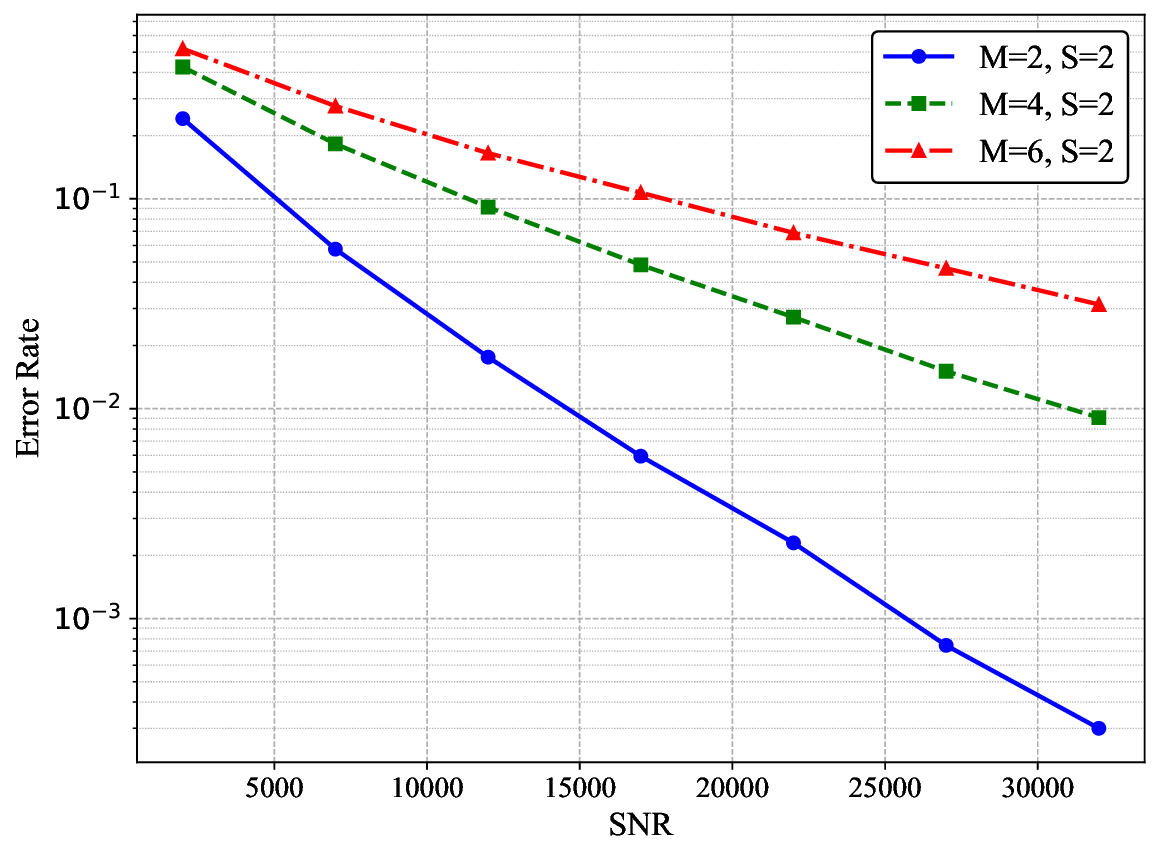} 
    \caption{Impact of transmitter count on error rate performance with varying SNR levels. }
    \label{fig:4}
\end{figure}

In Fig. \ref{fig:3}, the error rate is evaluated for a fixed number of transmitters \( M = 2 \) while varying the computational combinations \( S \). It can be observed that the error rate increases significantly as \( S \) grows, under the same signal-to-noise ratio (SNR) conditions. This is because a larger \( S \) introduces greater computational complexity, leading to increased molecular interactions and a higher likelihood of errors in the received signals.

At lower SNR levels, the error rate remains relatively high across all configurations due to the dominance of noise and molecular randomness. However, as the SNR improves, the error rate decreases steadily, and the performance gap between different computational combinations becomes more pronounced. Notably, the configuration with \( S = 2 \) achieves the lowest error rate, reflecting its reduced computational complexity and minimal molecular interference compared to higher \( S \) values.

These results emphasize the trade-off between computational complexity and system reliability. While larger \( S \) enables more complex operations within the molecular communication channel, it also introduces additional challenges for accurate computation, particularly in noisy environments. Therefore, optimizing the choice of \( S \) is critical for balancing computational performance and system robustness.

We now analyze the impact of the transmitter count \( M \) on the system's error rate performance, as shown in Fig. \ref{fig:4}. Contrary to expectations, the results demonstrate that increasing the number of transmitters does not always lead to improved system reliability. Specifically, as \( M \) increases, the error rate exhibits an upward trend across the entire range of SNR values.  

\begin{figure}[t]
    \centering
    \subfloat[Theoretical vs simulated error rates for \( M=4 \), \( S=2 \).]{
        \includegraphics[width=0.3\textwidth]{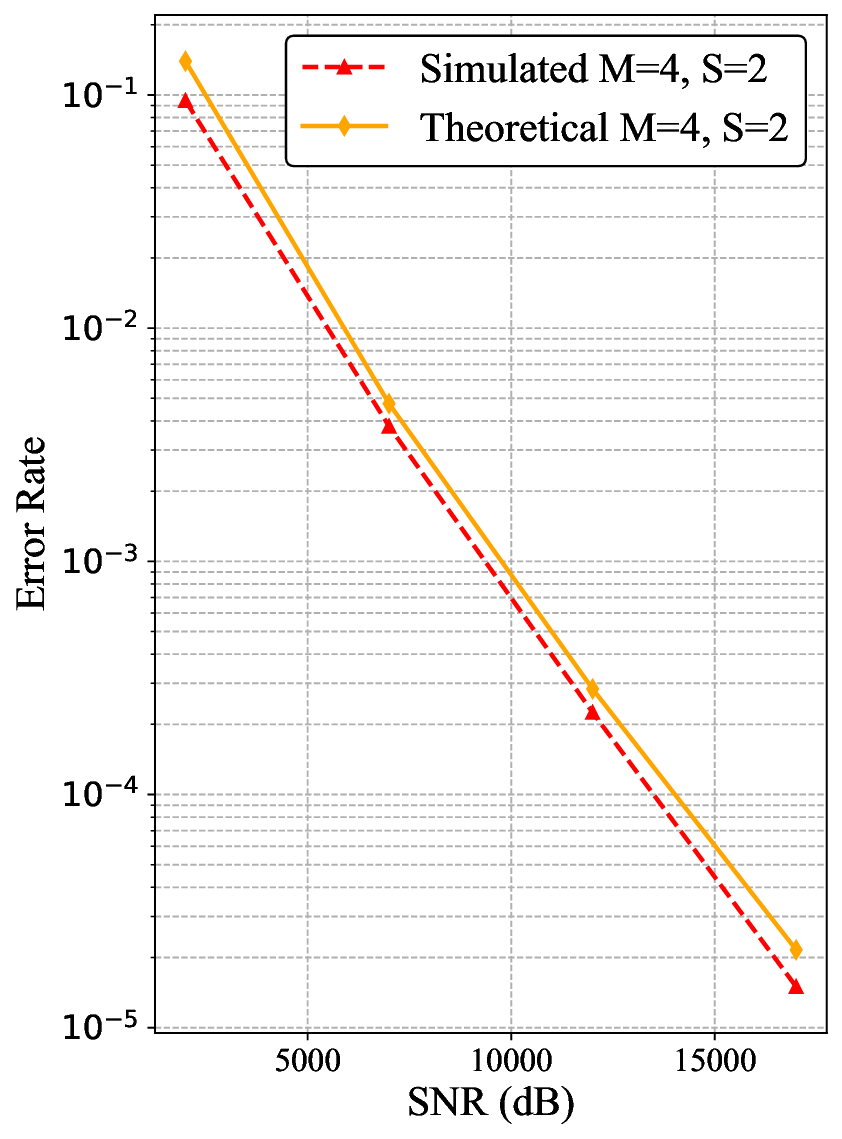}
        \label{fig:sub_a}
    }    

    \subfloat[Theoretical vs simulated error rates for \( M=4 \), \( S=4 \).]{
        \includegraphics[width=0.3\textwidth]{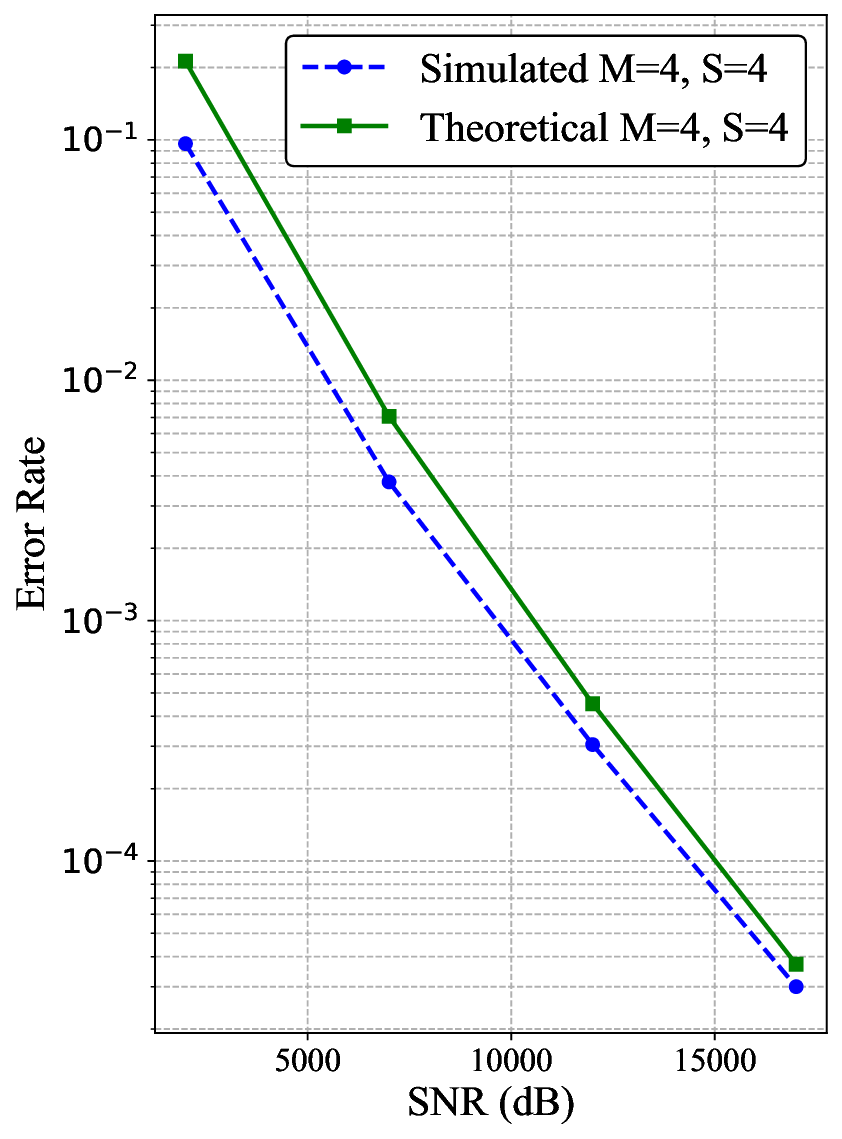}
        \label{fig:sub_b}
    }
    \caption{Comparison of theoretical and simulated error rates for different MC configurations.}
    \label{fig:cccc}
    
\end{figure}

This phenomenon can be attributed to several key factors. First, molecular interference becomes more pronounced as the number of transmitters increases. While the total number of molecules released into the environment grows proportionally, the superposition of molecular signals introduces challenges in accurately distinguishing signal types at the receiver. Second, the diffusion asymmetry between molecule types \( A \) and \( B \), with diffusion coefficients \( D_A = 2.2 \times 10^{-9} \, \text{m}^2/\text{s} \) and \( D_B = 1.5 \times 10^{-9} \, \text{m}^2/\text{s} \), respectively, exacerbates the signal dispersion, leading to increased uncertainty in molecular arrival times. 

In Fig. \ref{fig:cccc}, we compare the theoretical error rate upper bound with the simulated error rate under two configurations: \( M=4, S=2 \) with \( N=60 \) sampling points and \( M=4, S=4 \) with \( N=120 \) sampling points. The theoretical error rate is derived from the upper bound based on the probabilistic molecular communication model, while the simulated error rate is obtained through particle-based Monte Carlo simulations.

In Fig. \ref{fig:cccc}(a), the theoretical and simulated curves show reasonable alignment, though minor discrepancies arise due to the limited sampling density. In Fig. \ref{fig:cccc}(b), the alignment between theoretical predictions and simulation outcomes improves significantly, demonstrating the benefit of enhanced sampling points. Moreover, as the number of transmitted molecules increases, the theoretical and simulated curves exhibit even better agreement, reinforcing the robustness of the model. However, it is important to note that the simulations were not fully completed for higher molecular counts due to the prohibitive computational complexity. Despite this limitation, the trends observed in the figures clearly indicate that the theoretical and simulated results converge as the transmitted molecular count increases, further validating the accuracy of the proposed framework.

As the SNR increases, the effect of probability overlap diminishes. The noise becomes less dominant, and the molecular arrival probabilities corresponding to different hypotheses become more distinguishable. Consequently, the MAP detection method can more accurately identify the transmitted molecular signals, and the simulated error rate aligns closely with the theoretical upper bound.

By comparing the results of Fig. \ref{fig:cccc}(a) and Fig. \ref{fig:cccc}(b), we further observe the impact of the number of sampling points on system performance. With \( N=60 \) sampling points in Fig. \ref{fig:cccc}(a), the low sampling resolution introduces greater uncertainty, exacerbating the overlap in detection probabilities and increasing the likelihood of detection errors. In contrast, Fig. \ref{fig:cccc}(b) demonstrates that with \( N=120 \) sampling points, the higher sampling precision significantly reduces the probability overlap, leading to more accurate detection and improved alignment with the theoretical error rate.

\section{CONCLUSION}
This paper introduced a computational MC framework that seamlessly integrates arithmetic operations into the communication process, overcoming a fundamental limitation of traditional MC systems. By encoding positive and negative values with two types of molecules, the system achieves addition and subtraction through molecular interactions. An asymptotically tight BER upper bound was derived. Simulations validated the framework's robustness under molecular noise and environmental variability. These results establish the proposed framework as a significant step toward enabling intelligent and adaptive nanoscale systems, with promising applications in biochemical control, environmental sensing, and targeted drug delivery.

\section*{Acknowledgments}

Natalio Krasnogor acknowledges the UK’s Department for Science, Innovation and Technology (DSIT) and the Royal Academy of Engineering Chair in Emerging Technologies award.
\balance
\bibliographystyle{IEEEtran}
\bibliography{ref}

\end{document}